\documentclass[conference]{IEEEtran}
\IEEEoverridecommandlockouts

\usepackage{flushend}

\usepackage{cite}
\usepackage{amsmath,amssymb,amsfonts}
\usepackage{algorithmic}
\usepackage{graphicx}
\usepackage{textcomp}
\def\BibTeX{{\rm B\kern-.05em{\sc i\kern-.025em b}\kern-.08em
    T\kern-.1667em\lower.7ex\hbox{E}\kern-.125emX}}
\begin{document}

\newtheorem{mydef}{Definition}

\graphicspath{{./img/}}

\title{Authentication and Encryption for a Robotic Ad Hoc Network using Identity-Based Cryptography\\
}

\author{\IEEEauthorblockN{Jonay Su\'arez-Armas, Alexandra Rivero-Garc\'ia, Pino Caballero-Gil and C\'andido Caballero-Gil}
\IEEEauthorblockA{\textit{Department of Computer Engineering and Systems} \\
\textit{University of La Laguna}\\
Tenerife, Spain \\
\{jsuarear,ariverog,pcaballe,ccabgil\}@ull.edu.es}
}

\maketitle

\begin{abstract}
In some situations the communications of a place can be affected, totally lost, or not even exist. In these cases, the MANETs play an important role, allowing to establish a communications point using the different nodes of the network to reach the destination using decentralized communications. This paper proposes the implementation of a Robotic MANET, a decentralized network using robots as its nodes, which allows to move the network nodes to the desired location remotely. For this, each robot has as a core a Raspberry Pi with the capabilities to perform audio and video streaming, remote control of robots, tracking of objects, and deployment of wireless networks. To protect the network, different security mechanisms are used that allow secure authentication on the network by different nodes and encryption of information transmitted between them. All communications are protected through Identity-Based Cryptography, specifically with an Identity-Based Signcryption scheme.
\end{abstract}


\section{Introduction}\label{sec.introduction}
In a world in which technology has been incorporated, both in business and homes, communications have become essential. Due to the centralization of information and storage of data in the cloud, a failure in communications (especially in the business environment) can cause chaos. If this happens in a population nucleus instead of an enterprise, the consequences can be much more serious, especially if it is due to a natural disaster or some kind of catastrophe.

When there is a communications outage situation, it is about finding a solution, establishing a temporary communications point that allows, at least, to have a connection point that allows basic communications to maintain control of the situation.

\begin{figure}[htbp]
\centering
\includegraphics[width=0.45\textwidth]{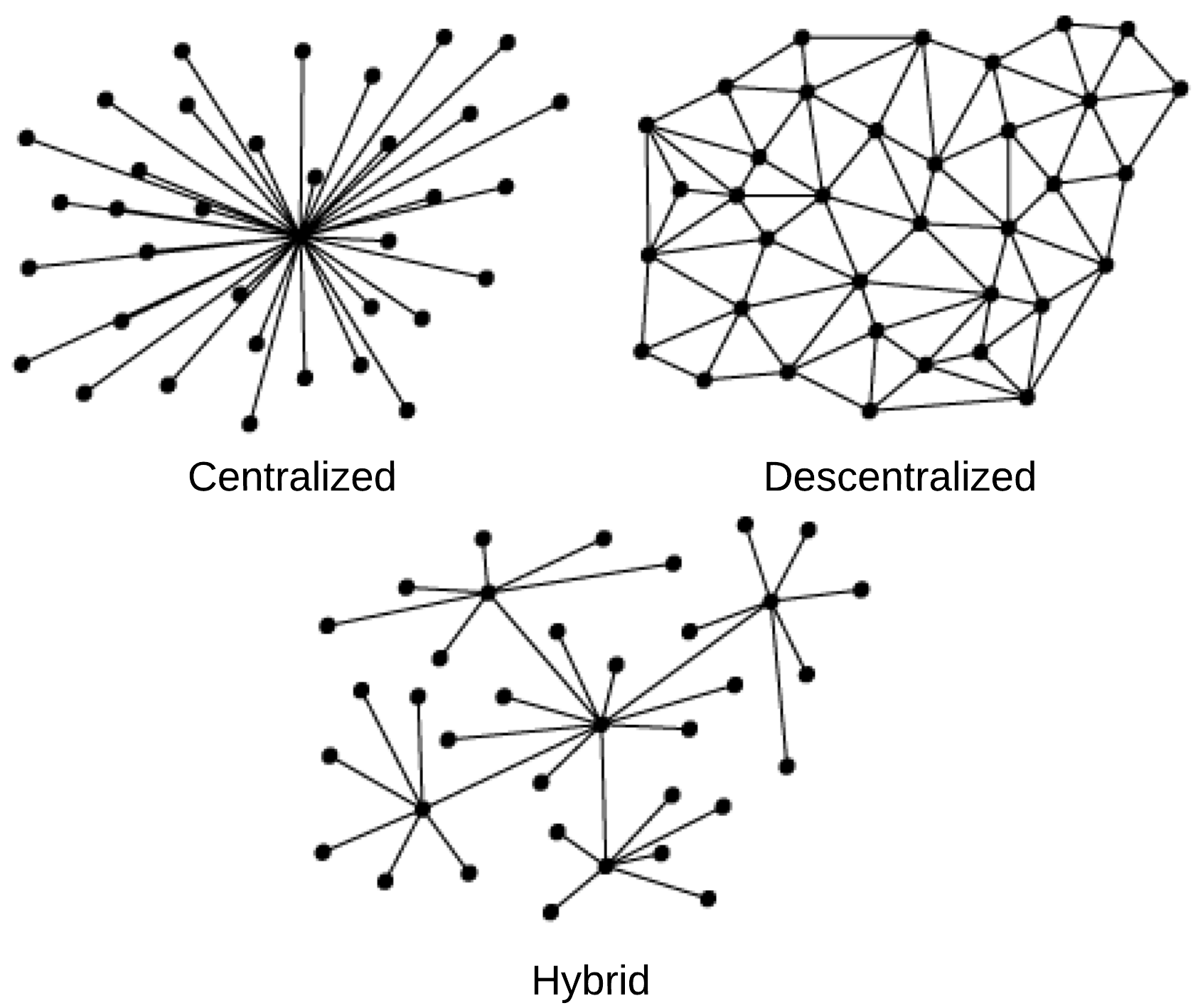}
\caption{Network topologies.}
\label{networktopologies}
\end{figure}

Unlike centralized communications, decentralized or distributed communications enable direct communication between devices in a network. This means that in this type of communications it is not necessary to have a central device so that the different devices in the network can communicate. Between these two topologies, there is also an intermediate case called hybrid (See Fig.~\ref{networktopologies}).

In order to establish a communications point in a place where communications have broken down or a temporary communication point is desired, this paper proposes a MANET (Mobile Ad Hoc Network) based on the use of robots that could adopt the acronym R-MANET (Robotic Mobile Ad Hoc Network) or directly RANET (Robotic Ad Hoc Network). The robots will be in charge of forming a multi-hop \cite{ref8} network to finally provide a communications point to the place where the network final nodes (robots) are established.

Robotics has increasingly gone into homes, to the point that we have kitchen robots \cite{ref20}, or robots that clean the floor \cite{ref19} for us. The latter, vacuum cleaners have several advantages such as the possibility of returning automatically to the battery charging station when necessary, so they are being used for other tasks that are not related to cleaning. The proposal therefore uses vacuum robots as network nodes.

In this work, the security of the proposed system is an essential component. We work in the form of authentication of the different network nodes, verifying that each of them are authorized by the network. In addition, a way of encrypting the data is proposed, so that communications are secure and the information will be protected from the sender to the receiver. An IDentity-Based Signcryption (IDBS) \cite{refIDBS} is used for communication confidentiality, authenticity and integrity, both among peers, and between the central station and robots.

The following section discusses a series of works close to the proposal made. Section 3 explains the proposed system in more detail and section 4 also details the components of each robot-based network node. Section 5 shows the security mechanisms used in the proposed system. Finally, a conclusion close the paper.

\section{Related works}\label{sec.relatedworks}
One of the fields in which decentralised communications can be very important is in IoT \cite{ref1} (Internet of Things). More and more devices are connected, and it may be really interesting that between them they form a Wireless Sensor Network (WSN). Many times, this type of network can be the solution for sending data to the cloud when the device does not have a direct Internet connection, but jumping between devices is able to send the information to one with an Internet connection available. In \cite{ref11}, different Multi-Hop routing protocols for Ad Hoc wireless networks are studied and compared. In this type of network, it is important to measure transmission rates between nodes, so that you can choose the optimal way to send information more quickly and efficiently \cite{ref12}.

In IoT, different works are based on the use of Raspberry Pi for data collection through sensors that form WSAN (Wireless Sensor and Actuator Network), which form a MANET between the different devices for data transmission between them \cite{ref9}. Another device used in these sensor networks is the Arduino microcontroller, as in \cite{ref10}, which form a network of sensors for temperature monitoring and to help air conditioning systems operate more efficiently.

The increase in computing capacity of smartphones has also made them useful for decentralized networking. Bluetooth is one of the technologies chosen to form MANET networks between mobile devices \cite{ref13}. These devices are also used to create VANETs that allow the exchange of information between vehicles on the road \cite{ref14}, so that they can share some useful information such as traffic status. The ability to obtain data through built-in sensors means that smartphones can be mounted on a device to move it to one location and obtain data \cite{ref15}, sometimes forming a MANET to deploy several devices at once.

MANET networks are also mounted on drones, such as in \cite{ref2}, where these networks are used to increase the coverage range of the remote control.

The iRobot Roomba robots apart from the tasks for which they have been designed \cite{ref5}, are able to perform other actions thanks to the Open Interface that it has, through which it is possible to send commands. This, in addition to the low cost of these robots, has caused them to be used for different jobs \cite{ref3}. In \cite{ref4}, the authors have worked on the control of an iRobot robot using an external keyboard and voice commands. In \cite{ref6}, a camera is placed on a Roomba robot to compare the captured video with one previously recorded to detect if there are abandoned objects in the examined area. Moreover, by applying fuzzy logic to these robots \cite{ref7} it is possible to develop navigation systems that allow them to obtain more optimal routes while are avoiding obstacles in their path.

In a decentralised network, it is very important to take into account its security. Two of the aspects to be studied are the form of authentication in the network and the protection of the data that is sent between the different nodes until reaching the final destination. Some authors propose models for authentication and auto-configuration of nodes in MANET networks \cite{ref16} and even on preventing impersonation attacks using double authentication factor \cite{ref17}. Apart from this, \cite{ref18} studies different cryptographic protocols for use in MANETs at the level of energy consumption, which is important in mobile devices, concluding that the AES standard is the one that produces the best performance among the different algorithms analyzed.

\section{Proposal description}\label{sec.proposaldescription}
The system proposed in this paper is a MANET in which the different network nodes are robots. The aim is to place the different nodes in the necessary positions to form a network in which a communications point is established at the end point. This communication point will provide the possibility of having a voice and/or data connection in an area that for some reason is out of communication, and where communications with other places are needed. The use of robots as network nodes allows each node to be moved remotely to the desired location, which can be important if it is a dangerous area for people to transport equipment.

\begin{figure}[htbp]
\centering
\includegraphics[width=0.489\textwidth]{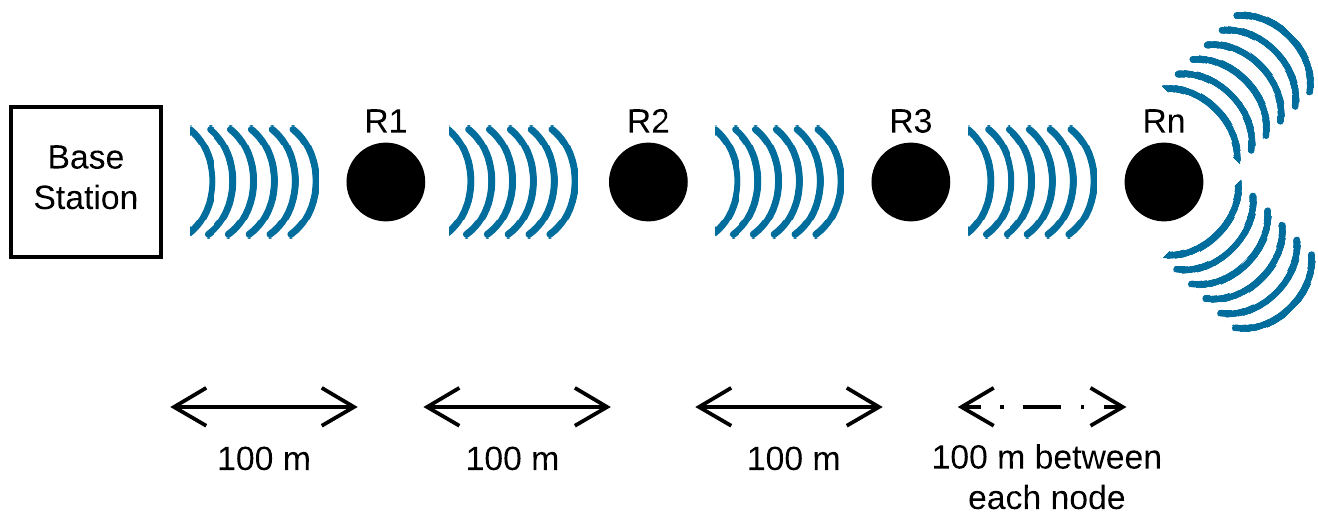}
\caption{Network scheme.}
\label{network}
\end{figure}

To start with the deployment of the robots that will form the network, first you must study the area and calculate the number of nodes needed to extend the network. This calculation is made considering the maximum range of wireless technology used for the connection between the nodes. It should be noted that in general, the theoretical maximum ranges in practice can decrease considerably. The technology chosen to deploy the network is Wi-Fi \cite{ref21}, which has a maximum range of 100 meters. If, for example, we wanted to establish a communications point 500 meters from the starting area, the theory tells us that we would have to form a network with 5 nodes to reach the desired area. To this we must add the base station located at the starting point, from which the signal is transmitted and the network is controlled. The diagram in Fig.~\ref{network} shows the general idea of the proposal.

The system proposed to deploy a temporary communication point through the use of robots is based on different functionalities: remote control of robots, streaming, object tracking and wireless network deployment.

First of all, there is the remote control of robots. This functionality allows each node of the network to be moved remotely to the point where it has been previously decided. From the base station it is possible to select a robot within the network and send it real-time movement commands.

\begin{figure}[htbp]
\centering
\includegraphics[width=0.489\textwidth]{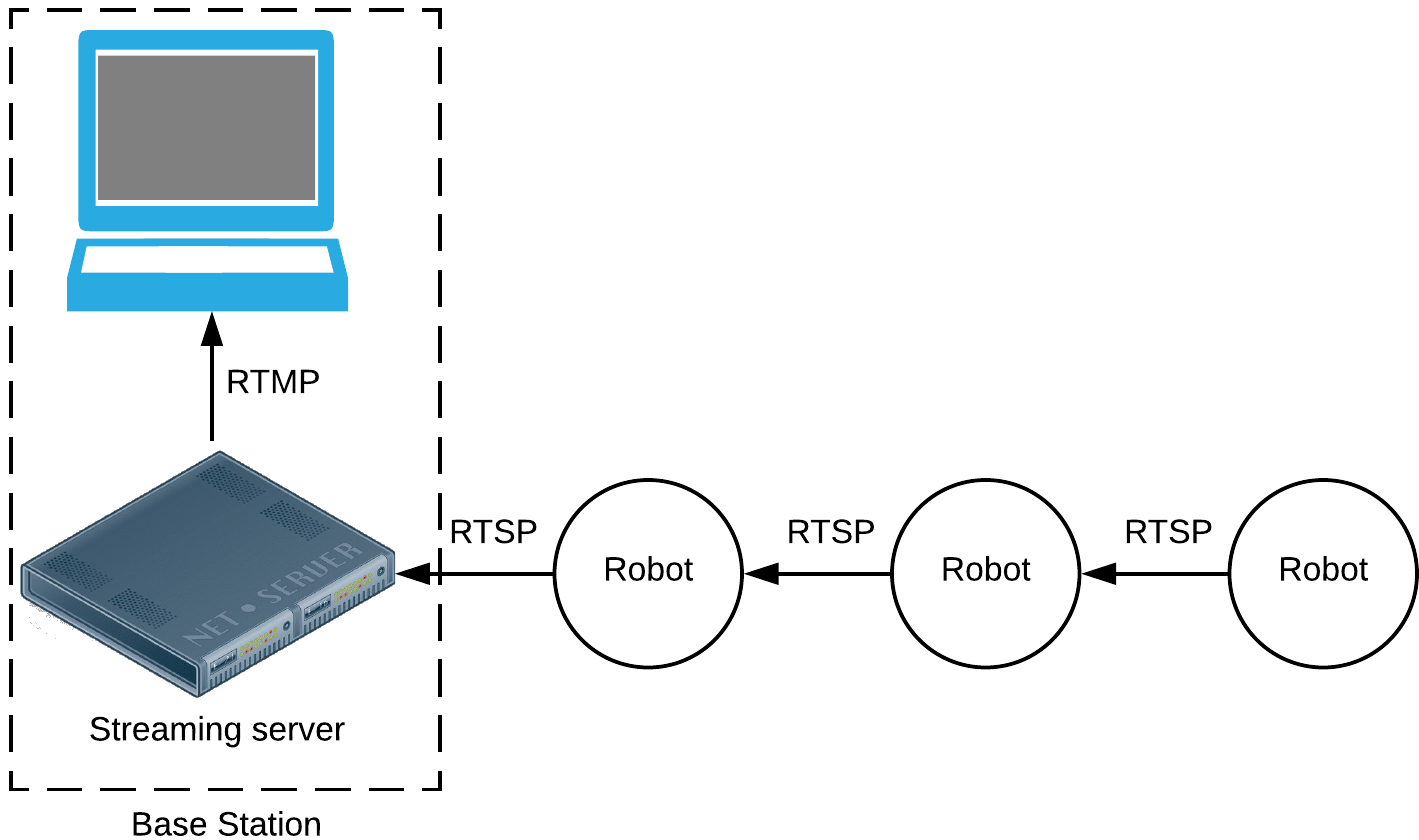}
\caption{Streaming scheme.}
\label{streaming}
\end{figure}

Secondly, we have audio and video streaming \cite{ref22}. To be able to move the robots remotely, it is necessary to know where they are at any given moment, so this second functionality is essential. Thanks to streaming, it is possible to visualize the environment in which the robot is located, at the same time that it is remotely controlled from the base station. In addition, audio is included in the streaming in case it would be useful to listen to the sound of the environment in which it is moving. To carry out this functionality, the base station must have a streaming server to which each robot will send the audio and video signal. The RTSP protocol \cite{ref23} is used to send this signal from the robots to the streaming server, and the RTMP protocol \cite{ref24} is used for receiving the video from the server to the base station equipment. In both cases, port 1935 is used (See Fig.~\ref{streaming}).

To perform these two functions, a web application is incorporated in the base station. It shows different blocks, and each one of the blocks refers to a robot, showing in them the video received through streaming together with buttons that allow sending movement orders to each robot.

\begin{figure}[htbp]
\centering
\includegraphics[width=0.25\textwidth]{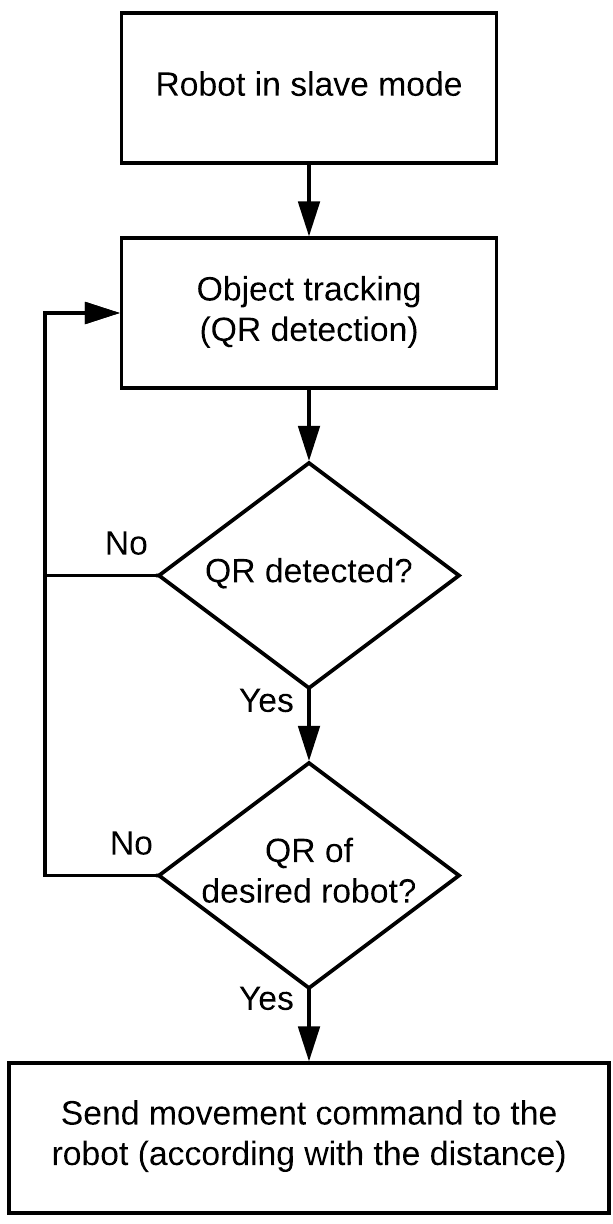}
\caption{Flowchart of robot tracking.}
\label{trackingflowchart}
\end{figure}

Third is object tracking \cite{ref25}, which is used for the most interesting part of the proposal. Directing each robot remotely to the desired location can be a tedious task (especially if the number of nodes is too large), so it is proposed that by controlling only one of the robots the rest will be able to follow it autonomously. For this, a master node will be selected and the rest will be slaves of it. Each robot has a QR code on the back that identifies it within the network. Each slave robot has to do image recognition, looking for any QR code in the images captured by your camera. This is when the robot has to check if the detected QR code corresponds with the robot you have programmed to follow. The flowchart of Fig.~\ref{trackingflowchart} shows this process.

\begin{figure}[htbp]
\centering
\includegraphics[width=0.48\textwidth]{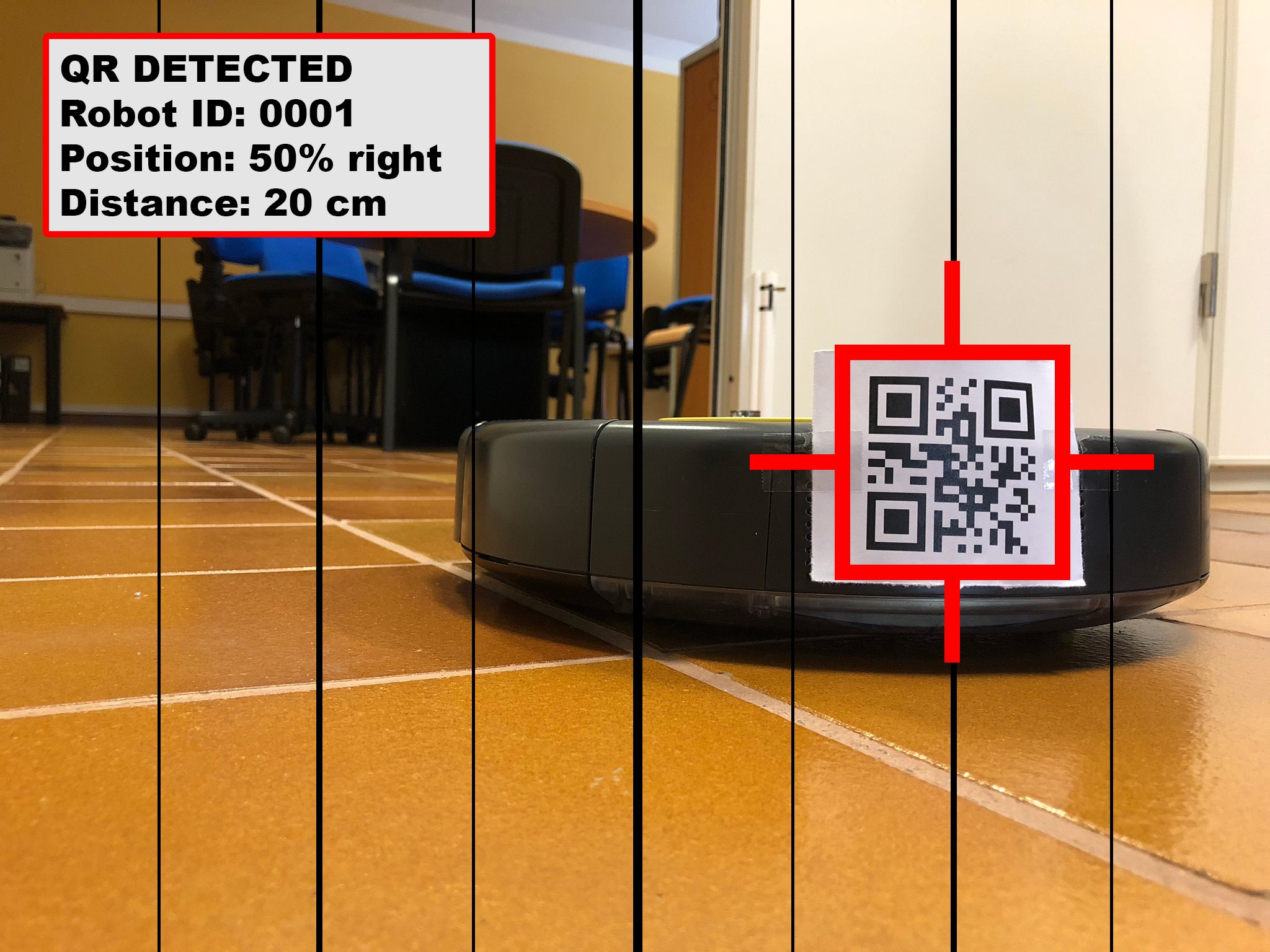}
\caption{QR detection.}
\label{qrdetection}
\end{figure}

Once a slave robot has detected the QR code of the robot to be followed, it performs calculations to establish the adequate movement. To do this, a proximity sensor is used, so that if the robot is more than a certain distance away, it has to start following the predecessor robot. In addition, depending on the area of the image in which the QR of the front node is detected, it will move in a certain direction (See Fig.~\ref{qrdetection}).

The latest functionality is responsible for deploying a wireless network at each node of the network. Each robot receives the Wi-Fi signal from its predecessor, so that it can have a connection to the base station by jumping between the different nodes in the network. In turn, they need to deploy a new Wi-Fi network so that the next node can connect to it to maintain a connection chain between the base station and the end node.

\section{Node description}\label{sec.nodedescription}
So far, the proposed system for deploying a RANET has been described, but the technologies and elements available to each network node have not been defined. It has also not been specified what type of robot is used.

Each node is built around an iRobot Roomba \cite{ref31} vacuum robot, which is connected to a Raspberry Pi, these two forming the core of the node. The Raspberry Pi is connected to the robot to send commands to it and is also responsible for establishing communications with the base station and the other nodes of the network.

\begin{figure}[htbp]
\centering
\includegraphics[width=0.489\textwidth]{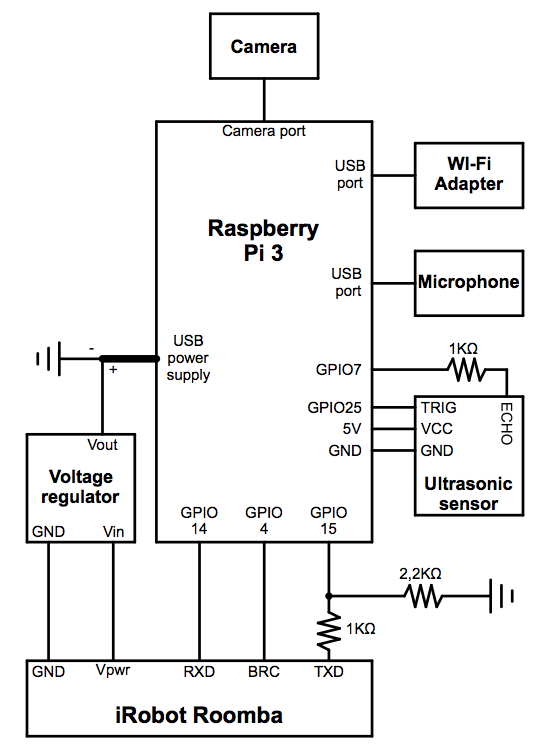}
\caption{Node wiring diagram.}
\label{node}
\end{figure}

As can be seen in the connection diagram of the different elements (See Fig.~\ref{node}), the Raspberry Pi is the central element of the node (all elements are connected to it thanks to the USB ports that it incorporates, GPIO pins or ports for specific use).

First of all there is the camera, which is connected through a specific port that incorporates the Raspberry Pi. Second are the USB ports, which are used to connect a second Wi-Fi adapter and a microphone for audio reception. Thirdly, an ultrasonic sensor is connected to the GPIO pins for detecting obstacles (adding a resistance that adjusts the tension provided on the Raspberry Pi) and the Roomba robot. To connect the robot, a resistive voltage divider \cite{ref32} is incorporated in the Roomba transmission pin to reduce the 5V supplied by it to a maximum of 3.3V, which allows the Raspberry Pi reception port.

$$ V_{out} = \frac{R_{2}}{R_{1} + R_{2}} \cdot V_{in} $$

And finally there is the connection of the Raspberry Pi's power supply. Taking advantage of the fact that the Roomba is able to provide power to external devices, the battery is used to avoid having to incorporate an external one, using a voltage regulator that converts the 14.4V that provides the Roomba battery to the 5V that Raspberry Pi needs.

In addition to the Wi-Fi interface that Raspberry Pi incorporates, a second interface has been added through the USB port to extend the network that gets the first interface, so that an interface is used in the usual reception mode, and the second is configured in Ad Hoc mode. For simpler configuration of network interfaces, as well as to allow external devices to connect to wireless networks created by different network nodes, a single class B network \cite{ref33} will be configured, which allows the connection of 65534 devices (discounting the base station and broadcast address), among which may be network nodes, or end devices. It should be noted that each network node will use 2 IP addresses, as it has two wireless cards; instead, each end device will use a single IP address. IP addressing of devices connected to the network is carried out automatically thanks to the incorporation of a DHCP server in the base station. To do this, a range of addresses is reserved for the network nodes, and all other addresses are assigned automatically.

\section{Security}\label{sec.security}
The security of communications is a crucial point of the system.
An ID-Based Signcryption scheme (IBSC) is used in order to achieve secure communications. 
This kind of cryptosystem is a combination of ID-Based Encryption (IBE) and ID-Based Signature (IBS), where all the shared messages are encrypted and signed. 
This proposed scheme offers the advantage of simplifying management of keys by not having to define a public key infrastructure.
A part of this, this scheme has a low computational complexity and it is efficient in terms of memory.
In this RANET system, the identification of nodes is a crucial point. 
Each robot has a QR code on the back that identifies it with a unique ID, this is the public identification used for the IBSC. 

First of all, some mathematical tools used is described. 
Let consider $(G, +)$ and $(V, \cdot)$ be two cyclic groups of the same prime order $q$.
$P$ is a generator of $G$ and there is a bilinear map paring $\hat{e} : G \times G \rightarrow V$ 
satisfying the following conditions:

\begin{itemize}
	\item Bilinear: $ \forall P, Q \in G$ and $\forall a, b \in \mathbb{Z}$, $\hat{e}(aP, bQ) 
= \hat{e}(P, Q)^{ab} $  
	\item Non-degenerate: $ \exists P_1, P_2 \in G $ that $\hat{e}(P_1,P_2) \neq 1$. This means if $P$ is generator of $G$, then $\hat{e}(P,P)$ is a generator of $Q$.
	\item Computability: there exists an algorithm to compute $\hat{e}(P,Q), \forall P,Q \in G$
\end{itemize}

\par Note, some hash functions denoted as follows are also needed: $H_1: \{0, 1\}^* \rightarrow G^*, H_2 : \{0, 1\}^* \rightarrow \mathbb{Z}^*_q, H_3 : \mathbb{Z}^*_q \rightarrow \{0, 1\}^n$,
where the size of the message is defined by $n$. 
Apart of this, $a \xleftarrow{r} N$ denotes a selection of an element $a$ randomly from a set $N$, 
$a \leftarrow b$ stands the assignation of the value $a$ to $b$ and $||$ is used for concatenation.   

The steps used in this signcryption scheme are the following: 
		
	\begin{itemize}
		\renewcommand{\labelitemi}{$\bullet$}
				
		\item \textbf{SETUP:} This step is execute in the base station to the establishment of the basic parameters and to the generation of the base station public key ($pk_{bs}$) and the base station secret key ($sk_{bs}$). To achieve this, a prime value $q$ is generated based on some private data $k \in \mathbb{Z}$ and the system select two groups $G$ and $V$ of order $q$ and a bilinear pairing map $\hat{e}: G \times G \rightarrow V$. $P \in G$ is selected randomly and the hash functions $H_1$, $H_2$ and $H_3$ are also chosen.   
		
$$ pk_{bs} \stackrel{r}{\leftarrow}  \mathbb{Z}^*_q $$
$$ sk_{bs} \leftarrow pk_{bs} \cdot P $$

		\item \textbf{EXTRACT ($ID_r$):} In this step, the keys for each robot is generated based on their ID.
		The robot $r$ with the identity $ID_r$ sends its information to the base station.
		The public key $Q_{ID_r} \in G$ and the secret key $S_{ID_r} \in G$ are calculated taking into account the $sk_{bs}$.
		This key exchange of data is performed in a secure way 
		
$$ Q_{ID_r} \leftarrow H_1(ID_r) $$
$$ S_{ID_r} \leftarrow  sk_{bs} \cdot Q_{ID_r} $$

		\item \textbf{SIGNCRYPTION ($S_{ID_{ra}}$, $ID_{rb}$, $m$):} If a robot $A$ with the identity $ID_{ra}$ shares a message $m \in\{0,1\}^n$ to the robot $B$ with the identity 
		$ID_{rb}$, the information will be encrypted and signed. The robot receiver's public key is generated taking into account the $ID_{rb}$ and then the message is signed with $S_{ID_{ra}}$ and encrypted with $Q_{ID_{rb}}$ giving as result $\sigma$ (a t-uple of three components: c, T, U).

$$ Q_{ID_{rb}} \leftarrow H_1(ID_{rb}) $$
$$ x \stackrel{r}{\leftarrow}\mathbb{Z}^*_q$$
$$ T \leftarrow x \cdot P$$
$$ r \leftarrow H_2(T || m)$$
$$ W \leftarrow x \cdot pk_{bs}$$
$$ U \leftarrow r \cdot S_{ID_{ra}} + W $$
$$ y \leftarrow \hat{e}(W, Q_{ID_{rb}}) $$
$$ k \leftarrow H_3(y)$$
$$ c \leftarrow k \oplus m$$
$$ \sigma \leftarrow (c, T, U)$$

	\item \textbf{UNSIGNCRYPTION ($ID_{ra}$, $S_{ID_{rb}}$, $\sigma$):} The robot $B$ receive $\sigma$ and parse the information.
	If everything is right, the message $m \in \{0,1\}^n$ is returned. Otherwise, if there are some problems in 
the signature or in the encryption of $m$, $\bot$ is returned. The sender's public key is 
generated taking into account $ID_{ra}$ and then the message is unencrypted with $S_{ID_{rb}}$.

$$ Q_{ID_{ra}} \leftarrow H_1(ID_{ra}) $$
$$ split \quad \sigma \quad as \quad (c, T, U) $$
$$ y \leftarrow \hat{e}(S_{ID_{rb}}, T)$$
$$ k \leftarrow y $$
$$ m \leftarrow k \oplus c $$
$$ r \leftarrow H_2(T || m) $$

Verification: 
$$\hat{e}(U, P) == \hat{e}(Q_{ID_{ra}}, pk_{bs} )^r \cdot \hat{e}(T, pk_{bs})$$
Note: if the verification is successful  $m$ is returned, otherwise $\bot$ is returned.

\end{itemize}

\section{Conclusion}\label{sec.conclusion}
MANET networks allow the connection between different devices and exchange information in a decentralized way. The behavior of such networks is similar to that of a P2P network. The main characteristic of this network is that the different nodes that form it are in motion. In a MANET network with robots, the ability to move nodes remotely is added. In this way it is not necessary to go to the different points where the nodes will be located, but it is possible to move them remotely thanks to streaming video visualization. In addition, the proposal incorporates object tracking, which allows that when moving one of the robots the rest moves behind the first one in an autonomous way.

Using Raspberry Pi as the primary device for each network node gives the system many possibilities. Having a device that runs a Linux operating system gives the possibility to add different services that run for some purpose and the development of applications that may be useful for system users. The USB ports and GPIO pins for device connection provide the possibility of adding different types of sensors for data acquisition or even actuators that can interact with the environment. The low battery consumption of these devices gives the possibility to use a small battery as a power source for these devices, as well as allowing a long use of them before draining the electricity provided by the battery.

In the proposed system, Wi-Fi is used as wireless technology in order to take advantage of the interface that incorporates Raspberry Pi, but it is possible to use any other type of wireless technology, allowing a greater range of coverage, which has an interface that can be connected to Raspberry Pi.

The incorporation of security mechanisms provides added value to the system, so that it can ensure that client devices connected to the generated network are secure. Using Identity-Based Cryptography we are able to provide the information encryption and authentication system by signing, with a low computational complexity.

Apart from the functionality for which this system has been designed, it can be used to perform video surveillance tasks using the camera and microphone incorporated in each of the network nodes, and can be moved thanks to robots to view images from different areas. This can be quite useful in hard-to-reach areas, or where people's physical integrity may be at risk.

To improve object tracking, in this case QR codes, in the future the design can be improved by including a small screen in the back of each robot. In this way, the permanent QR code is replaced by a screen that would allow the code to be changed from time to time. This change brings extra security to the system, since every few minutes the code used would become obsolete, thus preventing anyone from being able to move the robots to another location using a QR code that is too old.

\section*{Acknowledgments}
Research supported by the Spanish Ministry of Economy and Competitiveness, TESIS2015010102, the European FEDER Fund, and the CajaCanarias Foundation, under Projects TEC2014-54110-R, RTC-2014-1648-8, MTM2015-69138-REDT and DIG02-INSITU.

\end{document}